\documentclass[twocolumn,aps,pra,reprint,superscriptaddress,longbibliography]{revtex4-1}
\usepackage{amsmath}
\usepackage{graphicx}
\usepackage[lofdepth,lotdepth, caption=false]{subfig}
\usepackage{verbatim}
\usepackage{color}
\usepackage{hyperref}
\usepackage{dcolumn}
\usepackage{bm}
\usepackage{miller}
\usepackage{color}

\usepackage[english]{babel}

\begin{document}

\title{Evolution of the structural transition in Mo$_{1-x}$W$_{x}$Te$_{2}$}

\author{John A.~Schneeloch}
\affiliation{Department of Physics, University of Virginia, Charlottesville,
Virginia 22904, USA}

\author{Yu Tao}
\affiliation{Department of Physics, University of Virginia, Charlottesville,
Virginia 22904, USA}

\author{Chunruo Duan}
\altaffiliation[Present address: ]{Department of Physics and Astronomy, Rice University, Houston, Texas 77005, USA}
\affiliation{Department of Physics, University of Virginia, Charlottesville,
Virginia 22904, USA}

\author{Masaaki Matsuda}
\thanks{Notice:  This manuscript has been authored by UT-Battelle, LLC, under contract DE-AC05-00OR22725 with the US Department of Energy (DOE). The US government retains and the publisher, by accepting the article for publication, acknowledges that the US government retains a nonexclusive, paid-up, irrevocable, worldwide license to publish or reproduce the published form of this manuscript, or allow others to do so, for US government purposes. DOE will provide public access to these results of federally sponsored research in accordance with the DOE Public Access Plan (http://energy.gov/downloads/doe-public-access-plan).}
\affiliation{Neutron Scattering Division, Oak Ridge National Laboratory, Oak Ridge, Tennessee 37831, USA}

\author{Adam A.~Aczel}
\thanks{Notice:  This manuscript has been authored by UT-Battelle, LLC, under contract DE-AC05-00OR22725 with the US Department of Energy (DOE). The US government retains and the publisher, by accepting the article for publication, acknowledges that the US government retains a nonexclusive, paid-up, irrevocable, worldwide license to publish or reproduce the published form of this manuscript, or allow others to do so, for US government purposes. DOE will provide public access to these results of federally sponsored research in accordance with the DOE Public Access Plan (http://energy.gov/downloads/doe-public-access-plan).}
\affiliation{Neutron Scattering Division, Oak Ridge National Laboratory, Oak Ridge, Tennessee 37831, USA}
\affiliation{Department of Physics and Astronomy, University of Tennessee, Knoxville, Tennessee 37996, USA}

\author{Jaime A.\ Fernandez-Baca}
\thanks{Notice:  This manuscript has been authored by UT-Battelle, LLC, under contract DE-AC05-00OR22725 with the US Department of Energy (DOE). The US government retains and the publisher, by accepting the article for publication, acknowledges that the US government retains a nonexclusive, paid-up, irrevocable, worldwide license to publish or reproduce the published form of this manuscript, or allow others to do so, for US government purposes. DOE will provide public access to these results of federally sponsored research in accordance with the DOE Public Access Plan (http://energy.gov/downloads/doe-public-access-plan).}
\affiliation{Neutron Scattering Division, Oak Ridge National Laboratory, Oak Ridge, Tennessee 37831, USA}

\author{Guangyong Xu}
\affiliation{NIST Center for Neutron Research, National Institute of Standards and Technology, Gaithersburg, Maryland 20877, USA}

\author{J\"{o}rg C.\ Neuefeind}
\thanks{Notice:  This manuscript has been authored by UT-Battelle, LLC, under contract DE-AC05-00OR22725 with the US Department of Energy (DOE). The US government retains and the publisher, by accepting the article for publication, acknowledges that the US government retains a nonexclusive, paid-up, irrevocable, worldwide license to publish or reproduce the published form of this manuscript, or allow others to do so, for US government purposes. DOE will provide public access to these results of federally sponsored research in accordance with the DOE Public Access Plan (http://energy.gov/downloads/doe-public-access-plan).}
\affiliation{Neutron Scattering Division, Oak Ridge National Laboratory, Oak Ridge, Tennessee 37831, USA}

\author{Junjie Yang}
\altaffiliation[Present address: ]{Department of Physics, New Jersey Institute of Technology, University Heights, Newark, New Jersey 07102}
\affiliation{Department of Physics, University of Virginia, Charlottesville, Virginia 22904, USA}

\author{Despina Louca}
\thanks{Corresponding author}
\email{louca@virginia.edu}
\affiliation{Department of Physics, University of Virginia, Charlottesville,
Virginia 22904, USA}

\begin{abstract}
The composition dependence of the structural transition between the monoclinic 1T$^{\prime}$ and orthorhombic T$_{d}$ phases in the Mo$_{1-x}$W$_{x}$Te$_{2}$ Weyl semimetal was investigated by elastic neutron scattering on single crystals up to $x \approx 0.54$. First observed in MoTe$_{2}$, the transition from T$_{d}$ to 1T$^{\prime}$ is accompanied by an intermediate pseudo-orthorhombic phase, T$_{d}^{*}$. Upon doping with W, the T$_{d}^{*}$ phase vanishes by $x \approx 0.34$. Above this concentration, a phase coexistence behavior with both T$_{d}$ and 1T$^{\prime}$ is observed instead. The interlayer in-plane positioning parameter $\delta$, which relates to the 1T$^{\prime}$ $\beta$ angle, decreases with temperature as well as with W substitution, likely due to strong anharmonicity in the interlayer interactions. The temperature width of the phase coexistence remains almost constant up to $x \approx 0.54$, in contrast to the broadening reported under pressure.

\end{abstract}

\maketitle

\section{Introduction}

Quasi-two-dimensional van der Waals layered materials such as the transition metal dichalcogenides (TMDs) have received considerable attention recently because of their fascinating electronic properties arising from non-trivial band structure topologies, the emergence of superconductivity and its competition with a charge density wave instability, and extreme magnetoresistance. Weak interlayer coupling allows some layered materials to readily change their layer stacking and crystal symmetry by a sliding mechanism, often changing material properties as well. One such compound is the semimetal MoTe$_{2}$, in which a topology change accompanies layer stacking changes that occur on cooling.  
When MoTe$_{2}$ is cooled below $\sim$250 K, its layers slide from the stacking arrangement of the monoclinic 1T$^{\prime}$ phase, with the centrosymmetric space group $P2_1/m$, toward that of the orthorhombic T$_{d}$ phase, with the non-centrosymmetric space group $Pnm2_1$ \cite{clarke_low-temperature_1978}. The T$_{d}$ phase is reported to be a type-II Weyl semimetal \cite{sun_prediction_2015,deng_experimental_2016}. 
Recently, the 1T$^{\prime}$--T$_{d}$ transition was shown to exhibit surprisingly complex behavior, with transitions between both ordered and disordered layer stacking in various regions of the T$_{d}$--1T$^{\prime}$ thermal hysteresis loop \cite{tao_appearance_2019}. A new phase was found, the pseudo-orthorhombic T$_{d}^{*}$, which appears only on warming \cite{dissanayake_electronic_2019, tao_appearance_2019}. The kink in resistivity upon warming is likely due to the onset of T$_{d}^{*}$ \cite{tao_appearance_2019}. The tendency to form stacking superstructures has been reported in transmission electron microscopy measurements \cite{huang_polar_2019}. 

Despite the recent focus on van der Waals layered materials, transitions involving layer sliding, such as those found in MoTe$_{2}$, have not been studied extensively, especially when compared to other types of structural phase transitions such as those involving soft phonon modes \cite{a._cowley_structural_2006}. The ``sliding layer transition'' can be thought of as a distinct kind of structural phase transition, in which the layers shift from one stacking to another as a function of temperature or other external parameters. 
Materials with reversible temperature-induced sliding layer transitions include CrX$_{3}$ (X=Cl, Br, I) \cite{mcguire_crystal_2017} and $\alpha$-RuCl$_{3}$ \cite{glamazda_relation_2017}, of interest for their magnetic properties; Bi$_{4}$I$_{4}$ \cite{noguchi_weak_2019}, of interest as a weak topological insulator in its $\beta$ phase; and possibly Mn-doped CdPS$_{3}$ \cite{lifshitz_esr_1983}, though with conflicting evidence \cite{covino_synthesis_1985}. A better understanding of the conditions that allow sliding layer transitions to occur in MoTe$_{2}$ should help understand their presence in other materials as well. 

A multitude of factors besides temperature is known to influence the sliding layer transitions in MoTe$_{2}$, such as composition and pressure.
Increasing the W fraction as in Mo$_{1-x}$W$_{x}$Te$_{2}$ increases the transition temperature \cite{yan_composition_2017}. The effect of W substitution has been observed via transport measurements up to $x=0.57$ (but only up to 400 K) \cite{yan_composition_2017}, and inferred from room-temperature measurements throughout the phase diagram \cite{lv_composition_2017,oliver_structural_2017,rhodes_engineering_2017}, with WTe$_{2}$ long known to have the T$_{d}$ structure  \cite{brown_crystal_1966}. 
Other compositional variables include Te deficiency, which is reported to broaden the hysteresis in MoTe$_{2-y}$ crystals \cite{cho_te_2017}; Fe-doping, which lowers the transition by $\sim$80 K with $\sim$1\% impurities, broadening the transition while still preserving the asymmetry of the resistivity hysteresis loop \cite{huang_polar_2019}; and Nb substitution, which lowers and broadens the transition \cite{sakai_critical_2016}. 
Pressure decreases the transition temperature in MoTe$_{2}$ \cite{qi_superconductivity_2016, takahashi_anticorrelation_2017, lee_origin_2018, hu_angular_2019, heikes_mechanical_2018, dissanayake_electronic_2019}, with the T$_{d}$ phase not observed beyond $\sim$1.2 GPa. Pressure also broadens the temperature range of the transition, with the complex behavior at ambient pressure being replaced by a broad phase coexistence region where T$_{d}^{*}$ is no longer present \cite{dissanayake_electronic_2019}. 
In WTe$_{2}$, pressure also converts the T$_{d}$ phase to 1T$^{\prime}$, but only at a much higher pressure, with reported values of 4-5 GPa \cite{lu_origin_2016}, 6 GPa to 15.5 GPa \cite{zhou_pressure-induced_2016}, or 8 GPa \cite{xia_pressure-induced_2017}. 

Numerous other factors influence the transition in Mo$_{1-x}$W$_{x}$Te$_{2}$. Terahertz frequency light pulses are reported to drive WTe$_{2}$ to a centrosymmetric metastable state \cite{sie_ultrafast_2019}. In a thin film device, applied voltage can induce an electric field across WTe$_{2}$ that switches between different stacking orders \cite{fei_ferroelectric_2018, xiao_berry_2019}. An electron beam has been used to manipulate polar domain walls in MoTe$_{2}$ \cite{huang_polar_2019}. In-plane strain can subtly change the T$_{d}$--1T$^{\prime}$ transition temperatures in MoTe$_{2}$ \cite{yang_elastic_2017}. Non-hydrostatic pressure can preserve the T$_{d}$ phase of MoTe$_{2}$ beyond the point where hydrostatic pressure would render the T$_{d}$ phase extinct \cite{heikes_mechanical_2018}. Finally, reducing the out-of-plane thickness suppresses the transition in MoTe$_{2}$ flakes \cite{cao_barkhausen_2018,he_dimensionality-driven_2018,paul_controllable_2019}, typically with the T$_{d}$ phase remaining up to 300 K \cite{he_dimensionality-driven_2018}, though the atmosphere to which the flake is exposed may determine which phase is present \cite{paul_controllable_2019}. 

Despite the versatility in the ways of influencing the T$_{d}$--1T$^{\prime}$ transition, a detailed understanding of the transition mechanism is still lacking. A model based on free energy has been proposed \cite{kim_origins_2017} to explain the relative stability of the T$_{d}$ and 1T$^{\prime}$ phases. The lack of phonon softening as a function of pressure was reported to suggest that the T$_{d}$--1T$^{\prime}$ transition is entropy-driven \cite{heikes_mechanical_2018}. Meanwhile, a connection has been made to the Barkhausen effect \cite{cao_barkhausen_2018}, in which the magnetization of a ferromagnet exhibits sharp jumps when an applied magnetic field follows a hysteresis loop; for MoTe$_{2}$, temperature plays the role of field, and jumps in resistivity (due to changes in stacking) play the role of jumps in magnetization (due to individual magnetic domains switching their polarization). Each of these observations may provide essential details for a complete explanation of the transition.

\begin{figure}[h]
\begin{center}
\includegraphics[width=8.6cm]
{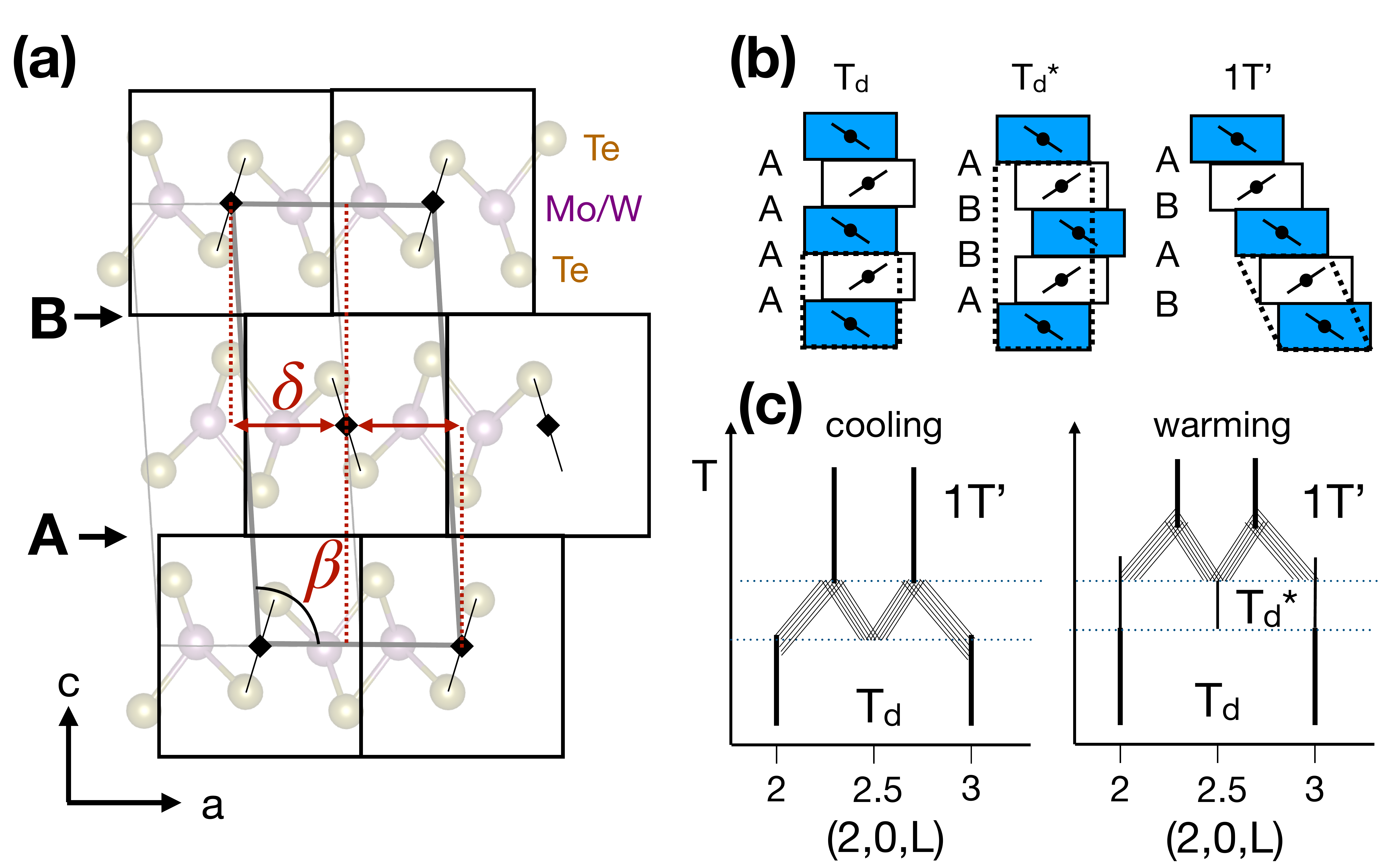}
\end{center}
\caption{(a) The crystal structure of 1T$^{\prime}$-Mo$_{1-x}$W$_{x}$Te$_{2}$ in the $a$-$c$ plane. Black diamonds indicate one set of inversion symmetry centers. (b) Stacking order for the T$_{d}$, T$_{d}^{*}$, and 1T$^{\prime}$ phases. Dashed-line boxes denote unit cell. (c) Schematic of changes in neutron scattering  intensity along $(2,0,L)$ on warming and cooling MoTe$_{2}$ through the transition.}
\label{fig:schematic2}
\end{figure}

The structure of  Mo$_{1-x}$W$_{x}$Te$_{2}$ can be captured in a picture of identical, centrosymmetric layers (Fig.\ \ref{fig:schematic2}(a)), stacked according to an A/B sequence of symmetry-equivalent stacking operations (Fig.\ \ref{fig:schematic2}(b)). These symmetries (identical and centrosymmetric layers) will be broken as the overall stacking requires, though only resulting in a small distortion of the atomic positions, e.g., $\lesssim$ $0.5$\% of the $a$ and $c$ lattice parameters between 1T$^{\prime}$ and T$_{d}$ as calculated from reported coordinates \cite{heikes_mechanical_2018}. 
In the ideal picture, these structures are specified by two parameters: the A/B stacking sequence; and the interlayer displacement parameter $\delta$. The $\delta$ parameter is ideally defined as the distance along the $a$-direction between the centers of inversion symmetry of neighboring layers as shown in Fig.\ \ref{fig:schematic2}(a), with $\delta$ chosen such that $0.5 \leq \delta \leq 1$. (More generally, we define $\delta$ as the distance between the midpoints of the metal-metal bonds, which coincides with the definition in the ideal case and for 1T$^{\prime}$, but also holds for the non-centrosymmetric T$_{d}$.)
Mathematically, if a layer can be characterized by a density $\rho(x,y,z)$, then the layer above would be ideally represented by  $\rho(-x + \delta, -y, z+1/2)$ or $\rho(-x + (1-\delta), -y, z+1/2)$ for A- or B-type stacking, respectively. 
The repeated stacking sequences for the phases are AA...\ for T$_{d}$, AB...\ for 1T$^{\prime}$, and AABB...\ for T$_{d}^{*}$. For their twins, these sequences are BB..., BA..., and ABBA..., respectively. In addition to these ordered phases, stacking disorder is present throughout the T$_{d}$--1T$^{\prime}$ hysteresis loop \cite{tao_appearance_2019,schneeloch_emergence_2019}, and analysis of neutron diffuse scattering suggests that this disorder is also consistent with an A/B stacking \cite{schneeloch_emergence_2019}. 

Irrespective of the stacking sequence, the positioning of atom pairs between adjacent layers is essentially the same.
In the ideal picture, an inversion operation applied to a structure that maps a particular layer onto itself will map surrounding layers from one stacking into another, reflecting the order of the A/B sequence of stacking operations while swapping every A with B and vice versa. Thus, A and B are, ideally, symmetry-equivalent stacking operations, and the parameter $\delta$ between any two adjacent layers would be the same regardless of A- or B-stacking. In reality, $\delta$ can still be uniquely defined for any adjacent layers in 1T$^{\prime}$ (derived from the $\beta$ angle and the $a$ and $c$ lattice parameters) or T$_{d}$ (derived from the atomic coordinates).  
From reported data in the literature, $\delta$ changes substantially with composition and less so with temperature. 
For example, values of $\delta$ in MoTe$_{2}$ for 1T$^{\prime}$ at 300 K and T$_{d}$ at 3 K, extracted from the neutron diffraction refinement data in Ref.\ \cite{heikes_mechanical_2018}, are 0.57386(10) and 0.5768(11), respectively, while $\delta = 0.56082(13)$ for WTe$_{2}$ at 113 K \cite{mar_metal-metal_1992}. 
Understanding trends in how $\delta$, the A/B stacking sequence, and the lattice parameters change as a function of temperature and composition could help clarify the conditions under which sliding layer transitions proceed.

In this paper, we report on elastic neutron scattering measurements on Mo$_{1-x}$W$_{x}$Te$_{2}$ single crystals, focusing on changes in the structure as a function of W substitution and temperature. The complex behavior seen in MoTe$_{2}$ is gradually replaced by a phase-coexistence-like behavior. By $x \approx 0.34$, the T$_{d}^{*}$ phase is no longer observed. The $\delta$ parameter decreases with both temperature and W-substitution. Increased W-substitution does not substantially change the transition temperature range up to $x\approx 0.54$, in contrast to the broadening seen under pressure.

\section{Experimental Details}
\label{sec:ExperimentalDetails}

The Mo$_{1-x}$W$_{x}$Te$_{2}$ crystals were grown via flux growth in excess Te. To begin with, Mo, W, and Te powders in some nominal ratio Mo$_{1-z}$W$_{z}$Te$_{2.1}$ were thoroughly mixed and pressed into a pellet. The pellet was sealed in an evacuated silica ampoule and sintered at 950 $^{\circ}$C for 1 to 2 days, followed by quenching in water. Next, the sintered material was pressed into pellets and sealed with Te flux in a typical mass ratio of 1:3 of Mo$_{1-z}$W$_{z}$Te$_{2.1}$ to Te in an evacuated silica ampoule, which was laid horizontally with a slight tilt downward toward the front of the box furnace. The ampoule was typically heated to 1150 $^{\circ}$C at a rate of 100 $^{\circ}$C/h, then steadily cooled down to 950 $^{\circ}$C in 1 to 3 weeks, followed by quenching in water or liquid nitrogen. 
(For $x \lesssim 0.08$, quenching is necessary to avoid the formation of the 2H phase \cite{rhodes_engineering_2017}.) The Te flux was subsequently removed by reheating the ampoule in a tube furnace at 900--1000 $^{\circ}$C, then decanting the liquid Te toward one end of the ampoule and quenching in water. 

The W fraction $x$ determined by energy-dispersive X-ray spectroscopy (EDS) was frequently much different from the nominal $x$, likely due to sample inhomogeneity. Let us define $c_{300\texttt{K}}$ as twice the interlayer spacing at 300 K (which coincides with the $c$ lattice parameter for the T$_{d}$ phase.) In Fig.\ S1 in the Supplemental Materials \cite{supplement}, we plot $c_{300\texttt{K}}$ (from X-ray diffraction) vs.\ $x$ (from EDS) for many small crystal flakes, showing plenty of scatter in $x$ but an overall nonlinear trend which we fitted with a quadratic curve. (See the Supplemental Materials for details \cite{supplement}.) The uncertainty in measuring $c_{300\texttt{K}}$ was small, as can be seen from the small scatter in $c_{300\texttt{K}}$ for the MoTe$_{2}$ and WTe$_{2}$ data sets (would would have no uncertainty in $x$.) Thus, in fitting the quadratic curve, we kept the curve fixed at mean $c_{300\texttt{K}}$ values at the MoTe$_{2}$ and WTe$_{2}$ endpoints. We used this relation to estimate $x$ for the crystals measured by triple-axis spectrometry from the $c_{300\texttt{K}}$ values obtained from the position of the $(004)$ peak. 
To account for uncertainty in measuring $c_{300\texttt{K}}$ between different crystals via triple-axis spectrometry, an uncertainty of $0.01$ \AA\ was assumed (in addition to uncertainties due to fitting), based on discrepancies observed in $c$ between different MoTe$_{2}$ crystals.
For MWT8, which was measured on the time-of-flight instrument CORELLI, $c_{300\texttt{K}}$ was obtained from the difference in the position of the $(004)$ and $(008)$ peaks.
Values of $c_{300\texttt{K}}$  are listed in Table \ref{tab:actualWfrac}. For crystals measured with neutron scattering, we used labels with ``MWT'' for crystals with $x > 0.01$, and ``MT'' for the remaining crystals. Crystal mass was typically on the order of $0.03$ g or higher, with the mass of the largest crystal, MWT1, being 0.6 g. We note that $x$ estimated from $c_{300\texttt{K}}$ is likely more relevant than $x$ estimated by EDS, since the neutron-scattering-estimated $x$ and the other neutron scattering measurements would be obtained from the same aligned crystallite.

\begin{table}[t]
\caption{Values of $c$ at 300 K extracted from neutron scattering data, and the effective W-fraction $x$ obtained from the relation described in the Supplemental Materials \cite{supplement}.}
\label{tab:actualWfrac}
\begin{ruledtabular}
\begin{tabular}{llll}
sample name & estimated $x$ & nominal $x$ & $c_{300\texttt{K}}$ (\AA)  \\
\hline
MWT1    & 0.087(25) & 0.1  & 13.810(10) \\
MWT2    & 0.171(25) & 0.4 & 13.842(10) \\
MWT3    & 0.21(4) & 0.2 & 13.855(13) \\
MWT4    & 0.335(25) & 0.3 & 13.897(10) \\
MWT5    & 0.420(25) & 0.5 & 13.922(10) \\
MWT6    & 0.505(25) & 0.7 & 13.946(10) \\
MWT7    & 0.542(25) & 0.7 & 13.955(10) \\
MWT8    & 0.445(6) & 0.5 & 13.9518(16) \\
\end{tabular}
\end{ruledtabular}
\end{table}

Some growth conditions differed from those listed above. For the MT2 crystal, a 2:1 mass ratio of Mo$_{0.8}$W$_{0.2}$Te$_{2}$ to Te flux was cooled from 1250 $^{\circ}$C to 950 $^{\circ}$C. Despite the nominal $x=0.2$ W-fraction, EDS and $c$-axis lattice constant measurements indicated a W-fraction of $x \lesssim 0.01$, so we labeled this crystal MT2. The other crystals labeled with ``MT'' are nominally $x=0$. For MT3, a Mo:Te mass ratio of 1:25 was used, and cooling was done from 1050 $^{\circ}$C to 900 $^{\circ}$C, as described in Ref.\ \cite{yang_elastic_2017}. The MT3 crystal was measured under pressure for the data in Ref.\ \cite{dissanayake_electronic_2019} before the ambient-pressure measurements in this paper. 
For the crystals measured by X-ray diffraction having $0.5 < x < 1$, the ampoule with the initial material was heated to 950 $^{\circ}$C at a rate of 100 $^{\circ}$C/h, then cooled to 750 $^{\circ}$C in 1 week. WTe$_{2}$ crystals were synthesized by sealing W and Te in a 1:13 molar ratio in an evacuated silica ampoule, and heating at 850 $^{\circ}$C in a tube furnace for 7 days. 

Powders of MoTe$_{2}$, Mo$_{0.8}$W$_{0.2}$Te$_{2}$, and WTe$_{2}$, used for neutron powder diffraction on the NOMAD instrument, were synthesized by sintering a stoichiometric mixture of powders of the elements, but with 5\% excess Te (e.g., Mo$_{1-z}$W$_{z}$Te$_{2.1}$.) After grinding, the powder was pressed into a pellet, sealed into an evacuated silica ampoule, and sintered at 950 $^{\circ}$C for MoTe$_{2}$ and Mo$_{0.8}$W$_{0.2}$Te$_{2}$, and 900 $^{\circ}$C for WTe$_{2}$, for at least 31 hours. Sintering was performed twice, with an intermediate mixing, followed by quenching in liquid nitrogen. Substantial decomposition of the WTe$_{2}$ powder occurred, with a $\sim$26\% weight fraction of elemental tungsten seen in refinement of neutron diffraction data, but no sign of Te deficiency was seen in refinement of the WTe$_{2}$ structure, which yielded a $2.03(3)$ Te occupancy.

Triple axis spectrometer measurements were performed on the HB1, CG4C, and HB1A instruments at the High Flux Isotope Reactor of Oak Ridge National Laboratory (ORNL), and on the SPINS instrument at the NIST Center for Neutron Research of the National Institute of Standards and Technology. All measurements were elastic, with incident neutron energies of 13.5 meV for HB1, 4.5 meV for CG4C, 14.6 meV for HB1A, and 5.0 meV for SPINS. The collimations for the HB1 and CG4C measurements were both 48$^{\prime}$-40$^{\prime}$-S-40$^{\prime}$-120$^{\prime}$, the collimation for HB1A was 40$^{\prime}$-40$^{\prime}$-S-40$^{\prime}$-80$^{\prime}$, and the collimation for SPINS was open-80$^{\prime}$-S-80$^{\prime}$-open. For SPINS, Be filters were used before and after the sample. For CG4C, a Be filter was used after the sample. 
Time-of-flight single crystal neutron diffraction measurements were performed on CORELLI \cite{ye_implementation_2018} at the Spallation Neutron Source (SNS) at Oak Ridge National Laboratory (ORNL). 

The NOMAD instrument \cite{neuefeind_nanoscale_2012} at the SNS of ORNL was used for the powder neutron diffraction measurements. For these measurements, 7 g of MoTe$_{2}$ and 2 g each of Mo$_{0.8}$W$_{0.2}$Te$_{2}$ and WTe$_{2}$ were sealed into vanadium cans and measured at a sequence of temperatures, on cooling from $\sim$300 K, for 3 hours per scan. Rietveld refinement was done using the \textsc{GSAS-II} software \cite{toby_gsas-ii:_2013}. 

All errorbars denote one standard deviation unless otherwise stated. (As noted above, estimates of $x$ for crystals used in triple-axis spectrometry measurements include an uncertainty of 0.01, which is an estimate of deviations in $c$ when measuring different crystals of the same composition.) 

For simplicity, all coordinates in this paper are reported in a unified orthorhombic coordinate system with $a \approx 6.3$ \AA, $b \approx 3.5$ \AA, and $c \approx 13.8$ \AA, with $c$ defined as twice the layer spacing, equivalent to the $c$-axis lattice constant in the T$_{d}$ phase.

\section{Data Analysis}
\label{sec:DataAnalysis}


\begin{figure}[h]
\begin{center}
\includegraphics[width=8.6cm]
{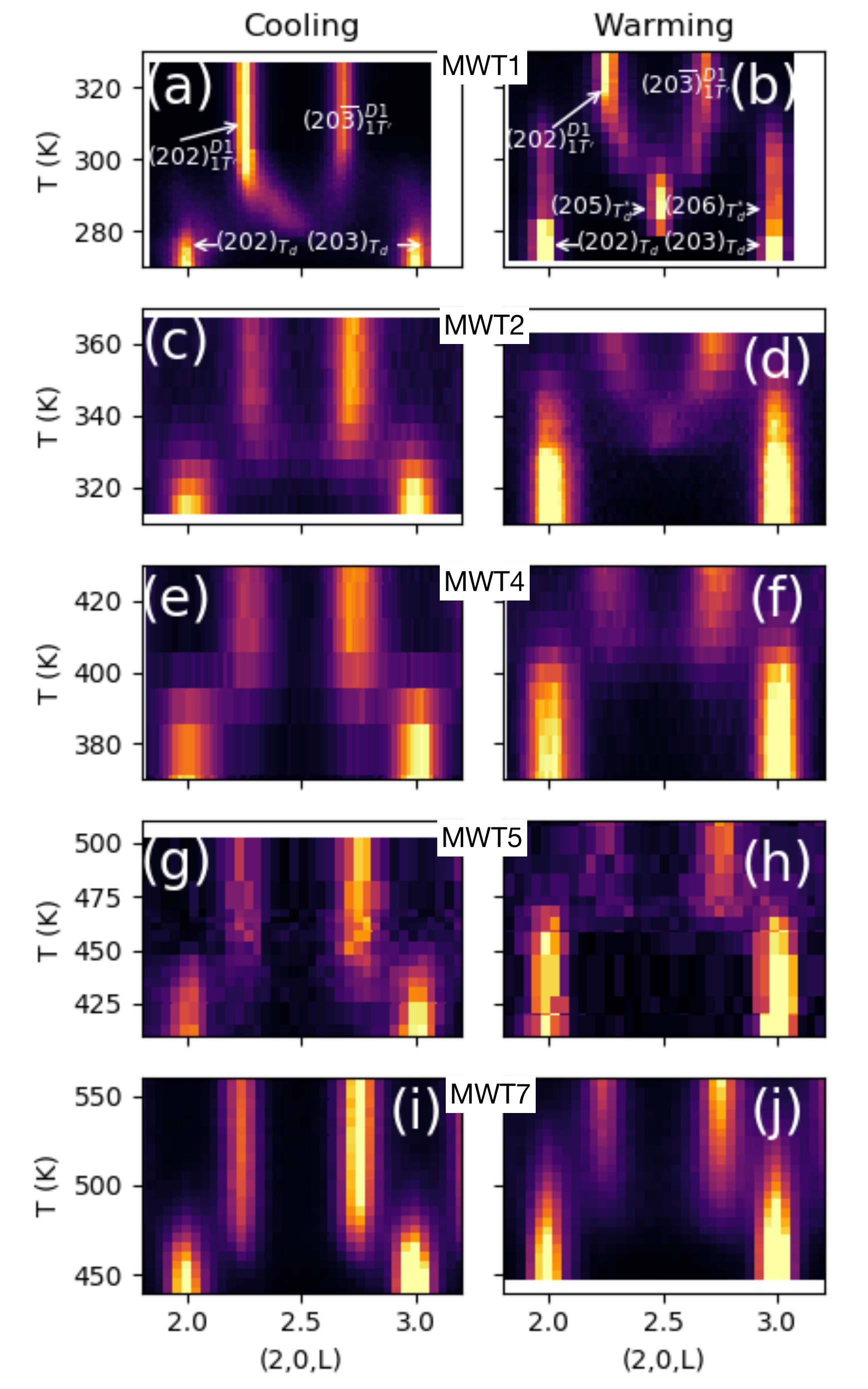}
\end{center}
\caption{Combined scans of neutron scattering intensity along (2,0,L) for many temperatures taken on (left) cooling and (right) warming. Data were taken on crystals (a,b) MWT1, (c,d) MWT2, (e,f) MWT4, (g,h) MWT5 and (i,j) MWT7, which are listed in order of increasing W-fraction as determined from $c_{300\texttt{K}}$. MWT1 was measured at CG4C, while the other four crystals were measured at HB1A. For MWT1, Bragg peaks are labeled. D1 and D2 refer to the two 1T$^{\prime}$ twins. For MWT1, the intensity jump at 300 K on cooling is due to combining data from two hysteresis loops.}
\label{fig:summary}
\end{figure}

Much information about the structural phase transitions in Mo$_{1-x}$W$_{x}$Te$_{2}$ can be obtained simply from the elastic neutron scattering intensity along $(2,0,L)$ \cite{tao_appearance_2019}, which is plotted as a function of temperature in Fig.\ \ref{fig:summary} for five crystals with different W-substitution levels. (Data on additional crystals are shown in the Supplemental Materials \cite{supplement}, including the exceptional cases of two nominally MoTe$_{2}$ crystals with very broad transitions \cite{supplement}.) The behavior for the lowest-doped crystal, MWT1 with $x \approx 0.09$, is similar to that of MoTe$_{2}$ \cite{tao_appearance_2019}.
Starting in the T$_{d}$ phase, we observe the $(202)_{T_{d}}$ and $(203)_{T_{d}}$ Bragg peaks at $L=2$ and $L=3$. On warming up to $\sim$280 K, the $(205)_{T_{d}^{*}}$ peak appears at $L=2.5$, indicating the arrival of the T$_{d}^{*}$ phase. Above $\sim$295 K, diffuse scattering appears in a ``V''-shape that spreads out from $L=2.5$ (and from $L=2$ and $L=3$ to some extent.) On further warming, MWT1 gradually transitions into 1T$^{\prime}$, and the $(202)_{1T^{\prime}}^{D1}$ and $(20\bar{3})_{1T^{\prime}}^{D2}$ Bragg peaks form near $L=2.3$ and $L=2.7$, where D1 and D2 denote each of the 1T$^{\prime}$ twins. On cooling, diffuse scattering appears in a similar V-shape as for warming, with intensity gathering toward $L=2.5$ but remaining diffuse. Then, the diffuse scattering diminishes, and the T$_{d}$ peaks at $L=2$ and $L=3$ are formed. 

With increasing W-fraction, there is a transformation from the complex transition of MoTe$_{2}$ to a simpler, phase-coexistence-like behavior. %
For MWT1 and MWT2 ($x \approx 0.17$), a T$_{d}^{*}$ peak at $L=2.5$ is seen on warming, but on cooling, the V-shaped diffuse scattering is much more subtle for MWT2 than MWT1. Similar behavior is seen in MWT3 with $x \approx 0.21$ as for MWT2 (see Supplemental Materials \cite{supplement}.) By MWT4 ($x \approx 0.34$), a T$_{d}^{*}$ peak is no longer clearly seen, though there is a subtle increase in intensity near $L=2.5$. Thus, the T$_{d}^{*}$ phase appears to be extinguished in the vicinity of $x \approx 0.2$ to $0.34$. By MWT5 ($x \approx 0.42$) and MWT7 ($x \approx 0.54$), only phase-coexistence-like behavior is seen. 

The V-shaped diffuse scattering can be explained by T$_{d}^{*}$-like regions appearing to varying degrees within an overall 1T$^{\prime}$ phase. From simulations of stacking disorder (see Fig.\ S6 in the Supplemental Materials \cite{supplement}), we see that coordinated shifts like A\textbf{BA}B$\rightarrow$A\textbf{AB}B result in the 1T$^{\prime}$ Bragg peaks near $L=2.3$ and $L=2.7$ moving closer to $L=2.5$. If T$_{d}$-like changes occurred instead (specifically, shifts such as ABA\textbf{B}AB$\rightarrow$ABA\textbf{A}AB that create and expand AAA... or BBB... regions), these 1T$^{\prime}$ peaks would instead move away from $L=2.5$, which is not seen in our data. Even when the movement of 1T$^{\prime}$ Bragg peaks along $(2,0,L)$ is small, the movement is always toward $L=2.5$. Thus, we see two patterns of behavior: diffuse scattering arising from T$_{d}^{*}$-like regions intimately mixed within an overall 1T$^{\prime}$ phase, along with an ordered T$_{d}^{*}$ phase present on warming; or phase coexistence of separated domains of T$_{d}$ and 1T$^{\prime}$.

\begin{figure}[h]
\begin{center}
\includegraphics[width=8.6cm]
{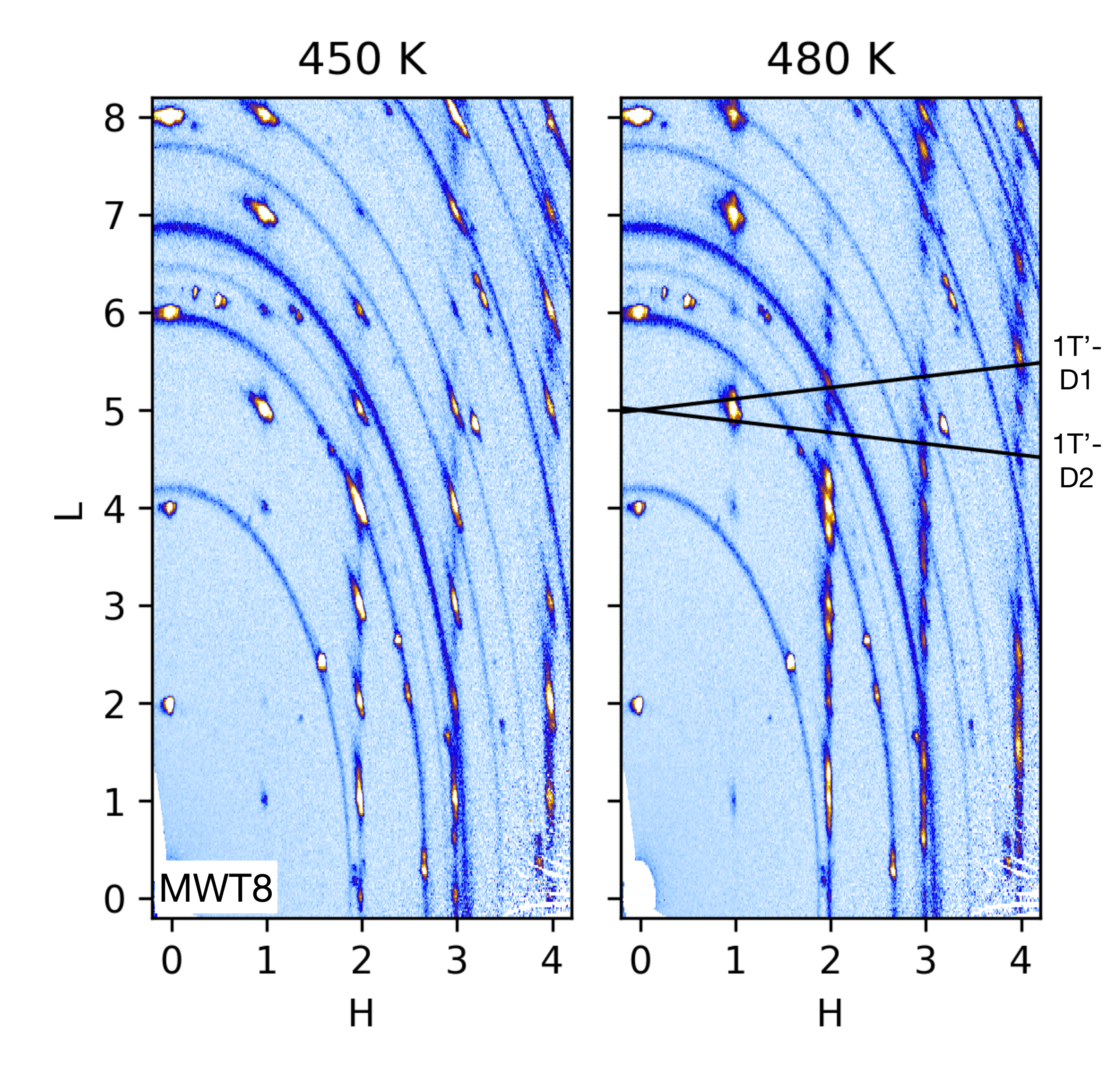}
\end{center}
\caption{Intensity maps of elastic neutron scattering in the H0L scattering plane on the MWT8 crystal with $x \approx 0.45$. Data taken at CORELLI at a sequence of temperatures on warming. To illustrate the divergence of the reciprocal space lattices of the two 1T$^{\prime}$ twins, at 480 K the $(H05)$ lines are plotted for each 1T$^{\prime}$ twin, labeled ``D1'' and ``D2''.}
\label{fig:CORELLI}
\end{figure}

The shift to phase-coexistence-like behavior is also seen beyond the $(2,0,L)$ line in reciprocal space. Fig.\ \ref{fig:CORELLI} shows intensity in the H0L plane for the MWT8 crystal (with $x \approx 0.45$) at two temperatures. Data were taken on CORELLI. At 450 K, the crystal is in the T$_{d}$ phase, but on warming to 480 K, additional 1T$^{\prime}$ peaks appear while the T$_{d}$ peaks are still present. No T$_{d}^{*}$ peaks at half-integer $L$ can be seen. (Additionally, our data show little change in the $0KL$ plane at different temperatures, consistent with layers sliding solely along the $a$-direction \cite{schneeloch_emergence_2019,tao_appearance_2019}. We observed no anomalies in the temperature-dependence of the $a$, $b$, and $c$ lattice parameters that were due to the transition.)

\begin{figure}[h]
\begin{center}
\includegraphics[width=8.6cm]
{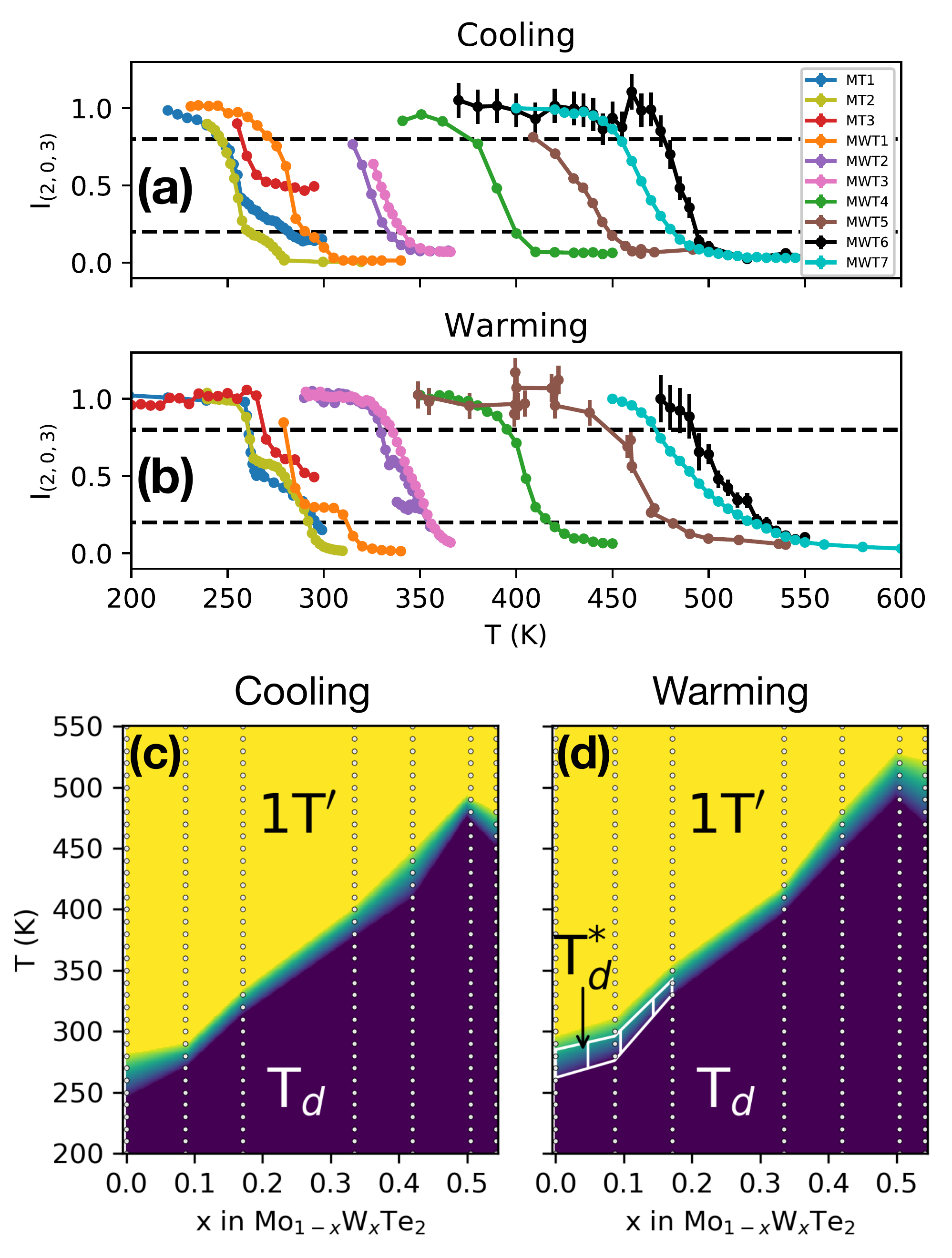}
\end{center}
\caption{(a,b) Neutron scattering intensity averaged near $(2,0,3)$ on (a) cooling and (b) warming. The plotted intensities $I_{(203)}$ are averages within $2.95 \leq L \leq 3.05$ of intensity along $(2,0,L)$, and normalized to the low-temperature (i.e., T$_{d}$ phase) value. 
The horizontal dashed lines show the intensities $I_{(203)} = 0.2$ and $I_{(203)} = 0.8$, used to define the beginning and completion temperatures of the transition.
The intensities for MT1 and MT2 are from the same data sets presented in Ref.\ \cite{tao_appearance_2019}, and intensities for most of the remaining crystals were obtained from data plotted in Fig.\ \ref{fig:summary} or Fig.\ S5 (in the Supplemental Materials \cite{supplement}.) For MWT1, a different data set (taken on HB1A) with a fuller hysteresis than that plotted in Fig.\ \ref{fig:summary} was used to obtain $I_{(203)}$. 
(c,d) Phase diagram plotting temperatures at which $I_{(203)}$ crosses 20\% and 80\% on warming and cooling. Dots denote where data were taken.
}
\label{fig:TvsC}
\end{figure}

To characterize how transition temperatures change as a function of W-fraction, we plotted $I_{(203)}$, the intensity averaged near $(2,0,3)$, as a function of temperature in Fig.\ \ref{fig:TvsC}(a,b). (A similar quantity, $I_{(2,0,2.5)}$, is shown in the Supplemental Materials in Fig.\ S7 \cite{supplement}.)
Generally, $I_{(203)}$ is maximum at low temperature due to the presence of the $(203)_{T_{d}}$ peak. At high temperatures, the crystal is in the 1T$^{\prime}$ phase, so no Bragg peaks are present near $(2,0,3)$ and $I_{(203)}$ is at a minimum. In between the extremes, two types of behaviors can be seen, depending on whether the transition is complex (with a T$_{d}^{*}$ phase and V-shaped diffuse scattering) or phase-coexistence-like. 

For crystals with lower W fraction (MT1, MT2, MT3, MWT1, and MWT2), pairs of ``onset'' temperatures can be seen, two on warming, and two on cooling. 
In MWT1, for example (Fig.\ \ref{fig:TvsC}(b)), on warming from T$_{d}$, the intensity drops abruptly at the onset to T$_{d}^{*}$ around $\sim$280 K. (We note that the transition to T$_{d}^{*}$ was already in progress when data collection started for the MWT1 data in Fig.\ \ref{fig:TvsC}(a,b).) The intensity then plateaus, and around $\sim$305 K a second onset temperature is passed, beyond which the gradual transition into 1T$^{\prime}$ starts. 
On cooling from 1T$^{\prime}$ (Fig.\ \ref{fig:TvsC}(a)), we again see two kinks in $I_{(203)}$, first at the onset into a frustrated T$_{d}^{*}$ region (with the V-shaped diffuse scattering) near $\sim$305 K, and then at the onset to T$_{d}$ near $\sim$280 K. The pairs of onset temperatures are close together on warming vs.\ cooling, despite the crystal entering and leaving different structures (as discussed for MT2 \cite{tao_appearance_2019}.) 

In contrast, for the higher W-substituted crystals (MWT4, MWT5, MWT6, and MWT7), the onset to the transition is less sharply defined, and no intermediate onset temperatures are evident. The onset temperatures for warming vs.\ cooling are not as close together as for lower W; for example, in MWT7, the transition begins around $\sim$480 K on cooling (into T$_{d}$) vs.\ 460 K on warming (into 1T$^{\prime}$.)

Fig.\ \ref{fig:TvsC}(c) shows a phase diagram of the transition as a function of the W fraction $x$. The upper and lower boundaries of the transition in the phase diagram denote temperatures at which 20\% or 80\% of $I_{(203)}$ is crossed. Transition temperature increases roughly linearly with $x$ up to $x \approx 0.5$. (The peak near $x=0.5$, as well as the similar anomaly for the $\beta$ angle in data presented below, is likely due to the estimated W fraction from $c_{300\texttt{K}}$ being slightly off, mis-ordering the true W fractions for the two $x \approx 0.5$ crystals.) 
The temperature range of the transition does not change substantially up to $x \approx 0.5$ when compared to the broadening seen under pressure \cite{dissanayake_electronic_2019}. 
However, though for $x<0.5$ the temperature range is about the same for warming and cooling, the range is noticeably broader on warming for MWT6 and MWT7 with $x \sim 0.5$. Interestingly, the transition under pressure also tends to become broader on warming \cite{dissanayake_electronic_2019}.


\begin{figure}[h]
\begin{center}
\includegraphics[width=8.6cm]
{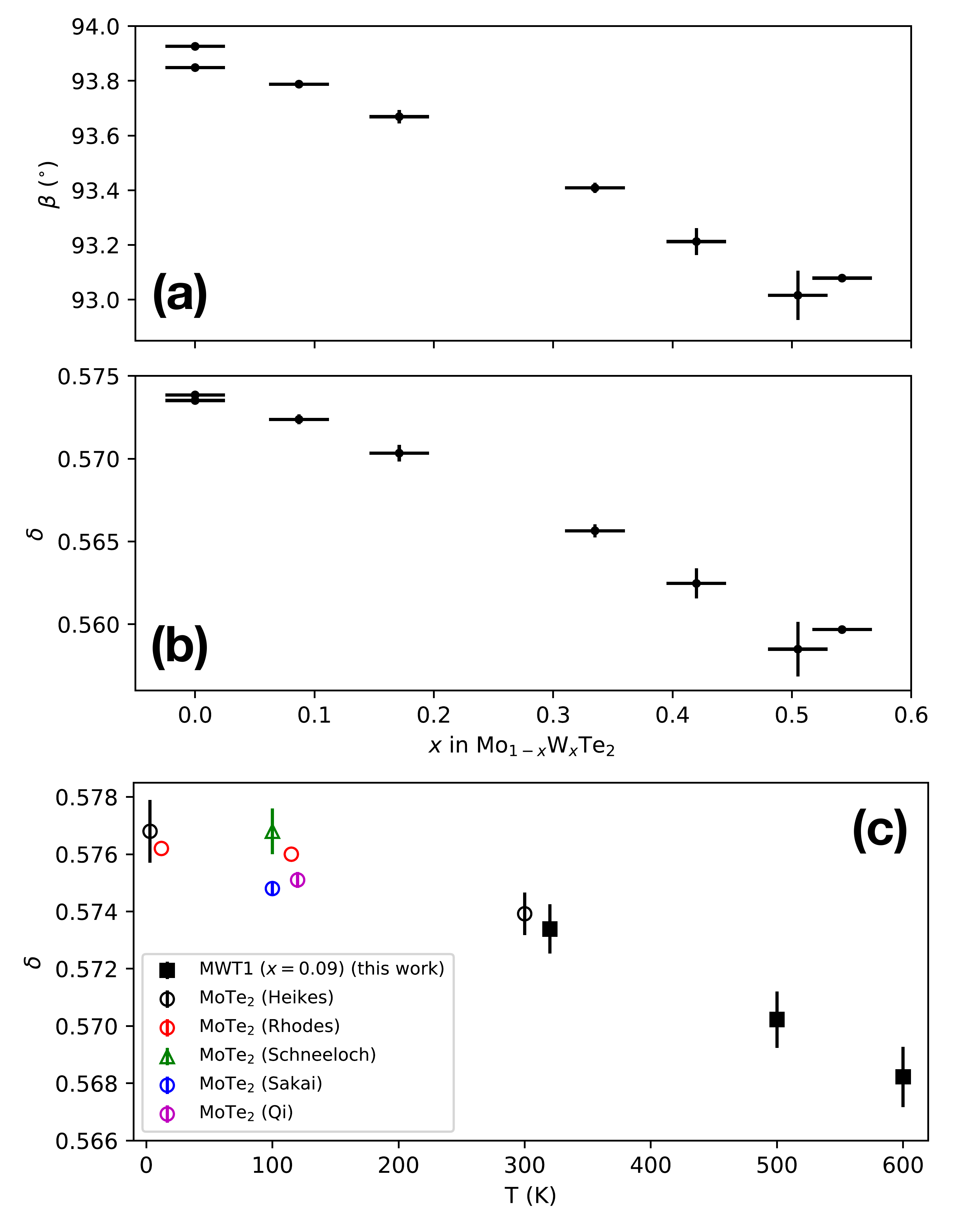}
\end{center}
\caption{(a,b) Changes in (a) 1T$^{\prime}$ $\beta$ angle and (b) interlayer displacement parameter $\delta$ plotted against $x$. The $\beta$ angle and $\delta$ were calculated from the positions of 1T$^{\prime}$ Bragg peaks along $(2,0,L)$ in single crystal data. 
(c) The $\delta$ parameter plotted as a function of temperature. The three MWT1 $\delta$ values were obtained from the separation of monoclinic Bragg peaks for MWT1 from data taken on SPINS. The remaining points are extracted from coordinates reported in the literature for the T$_{d}$ \cite{heikes_mechanical_2018, rhodes_bulk_2017, schneeloch_emergence_2019, sakai_critical_2016, qi_superconductivity_2016} (below 300 K) and 1T$^{\prime}$  \cite{heikes_mechanical_2018} (at 300 K) phases, labeled by first author.}
\label{fig:allBeta}
\end{figure}

We now shift our attention to the $\delta$ parameter, which gives information about the interactions between neighboring layers. The total energy is expected to have two minima with respect to layer displacements along the $a$-direction (as calculated for the interlayer shear optic mode of T$_{d}$ \cite{heikes_mechanical_2018}), with a separation that we label $\epsilon$. The $\delta$ parameter is related to $\epsilon$ via $1 + \epsilon = 2 \delta$. This parameter can be measured from the distance $s = 2 H \epsilon$ along $(2,0,L)$ separating neighboring Bragg peaks of opposite 1T$^{\prime}$ twins, if these twin peaks are centered about an integer $L$ location. The $\epsilon$ parameter (and thus $\delta$) is related to the 1T$^{\prime}$ monoclinic $\beta$ angle via $\sin(\beta - 90^{\circ}) = \frac{\epsilon a}{c}$ (where $a$ and $c$ refer to the 1T$^{\prime}$ lattice parameters.)

In Fig.\ \ref{fig:allBeta}(a,b) we plot the monoclinic $\beta$ angle and the $\delta$ parameter as a function of W fraction. We see the expected decrease in $\beta$ and $\delta$ with W-substitution \cite{oliver_structural_2017}. However, the $\delta$ parameter (and $\beta$) appears to have a significant temperature-dependence, as is seen in Fig.\ \ref{fig:allBeta}(c), where we plot $\delta$ for the MWT1 crystal between 320 and 600 K. We also plot $\delta$ extracted from MoTe$_{2}$ coordinates reported in the literature for the T$_{d}$ \cite{heikes_mechanical_2018, rhodes_bulk_2017, schneeloch_emergence_2019, sakai_critical_2016, qi_superconductivity_2016} and 1T$^{\prime}$ \cite{heikes_mechanical_2018} phases. The literature MoTe$_{2}$ $\delta$ values are consistent with the overall trend implied by our MWT1 $\delta$ values, though insufficient to determine the extent of any discontinuity in $\delta$ across the transition. 
The temperature-dependence of $\delta$ must skew the $\delta$ and $\beta$ values in Fig.\ \ref{fig:allBeta}(a,b) downward with increasing $x$, since these values were measured at temperatures that increased with W substitution (in accordance with the increasing transition temperature.) However, the decrease in $\delta$ with $x$ likely cannot be explained by the temperature-dependence alone, judging from the decrease in $\delta$ from 320 to 600 K in MWT1 
being only a fraction of the decrease in $\delta$ from $x=0$ to $x=0.5$ in Fig.\ \ref{fig:allBeta}(b).

The strong anharmonicity in vibrations involving layer displacement along the $a$-direction \cite{heikes_mechanical_2018} provides an explanation for the decrease in $\delta$ with temperature. If we assume two layers oscillate relative to each other  within a double-well potential (with each minimum corresponding to A- or B-type stacking), and the potential slope is less steep in directions toward the midpoint between the minima, then larger vibrations would be expected to shift the average interlayer position closer together, which corresponds to smaller (effective) $\epsilon$ and $\delta$. 
In principle, the data suggest a trend where, at a sufficiently high temperature, $\delta$ would approach 0.5 and $\epsilon$ would approach 0, and a higher-symmetry structure would result (dubbed T$_{0}$ and calculated to be unstable \cite{huang_polar_2019}); however, the rate of change of $\delta$ ($\sim$0.006 from 320 to 600 K) is too low for such a phase to be reached before thermal decomposition.

\section{Discussion}
\label{sec:Discussion}

\begin{figure}[h]
\begin{center}
\includegraphics[width=8.6cm]
{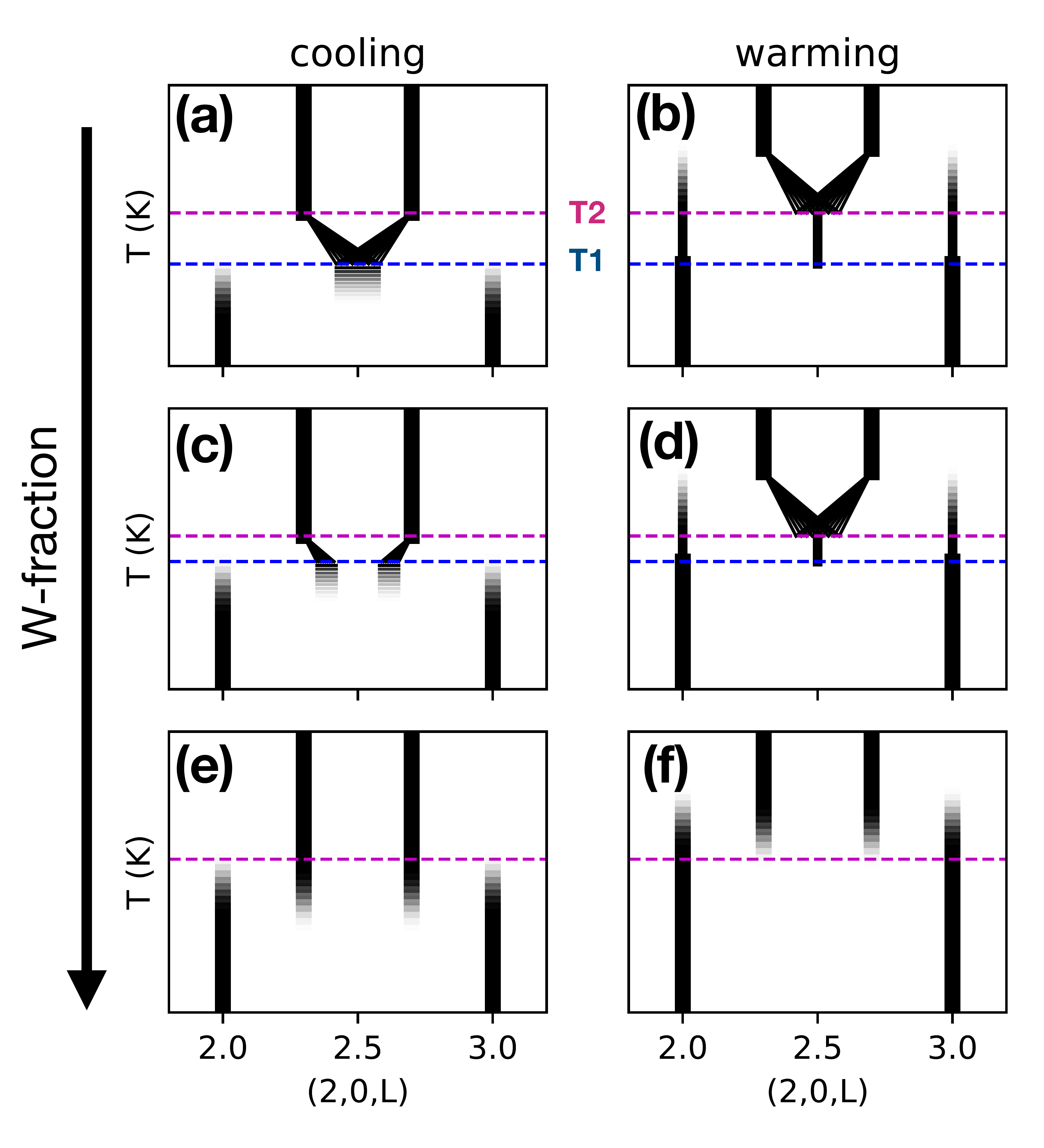}
\end{center}
\caption{Schematic diagram of transitions on (a,c,e) cooling and (b,d,f) warming for three levels of W-substitution: (a,b) $x = 0$, (c,d) $x \sim 0.2$, and (e,f) $x \gtrsim 0.4$.}
\label{fig:transition}
\end{figure}

To a large extent, Mo$_{1-x}$W$_{x}$Te$_{2}$ can be thought of as a modular material \cite{ferraris_crystallography_2008}, where the modules (the layers) can be rearranged locally in multiple ways, though unlike many modular materials, the rearrangement occurs reversibly with temperature. The small change in the interlayer spacing across the transition (as seen in Fig.\ S2 in the Supplemental Materials \cite{supplement}) highlights the similarity in the positioning of neighboring layers. 
Nevertheless, despite the near-symmetry-equivalence of the positioning of adjacent layers in Mo$_{1-x}$W$_{x}$Te$_{2}$, certain stackings are clearly preferred over others, depending on temperature and thermal history. 
For MoTe$_{2}$, we have argued \cite{tao_appearance_2019} that the onset temperatures discussed above provide boundaries for temperature ranges where certain types of short-range stacking are preferred. This situation is depicted in Fig.\ \ref{fig:transition}(a,b), with the onset temperatures marked $T1$ and $T2$, and assumed, for simplicity, to be the same on cooling vs.\ warming (though there is a $\sim$5 K hysteresis for the onset to or from T$_{d}$ \cite{tao_appearance_2019}.) For $T > T2$, ABABAB...\ stacking is preferred. For $T < T1$, AAAA...\ or BBBB...\ stacking is preferred. For $T1 < T < T2$, AABBAABB... stacking is preferred. However, the transition is often frustrated, and only the T$_{d}$$\rightarrow$T$_{d}^{*}$ and T$_{d}^{*}$$\rightarrow$T$_{d}$ transitions proceed without disorder. The onset temperatures only mark the \emph{beginning} of a transition toward a certain type of order; e.g., warming well above $T2$ is needed to transition essentially completely into 1T$^{\prime}$.

On W substitution, we propose that the $T1$ and $T2$ onset temperatures move closer together (Fig.\ \ref{fig:transition}(c,d)) and eventually merge (Fig.\ \ref{fig:transition}(e,f)) as the transition behavior becomes more phase-coexistence-like. On cooling, the V-shaped diffuse scattering becomes abbreviated, with a slight shift of intensity toward $L=2.5$ (as seen for MWT2 in Fig.\ \ref{fig:summary}(c).) On warming, the T$_{d}^{*}$ temperature range appears to shrink (Fig.\ \ref{fig:TvsC}(b)), though there is some ambiguity in estimating this range. Regardless, in our picture, we would expect moving $T1$ and $T2$ together to shrink the T$_{d}^{*}$ range, but not to shrink the V-shaped diffuse scattering until T$_{d}^{*}$ is completely gone, which would explain the presence of the V-shaped diffuse scattering on warming for MWT2 but not cooling (Fig.\ \ref{fig:summary}(c,d).) In this way, W substitution would not just increase the overall temperature of the transition, but move each onset temperature separately in ways that control which types of stacking are possible. 

As for the microscopic origin of these layer shifts, unfortunately, little theoretical analysis of the transition has been attempted beyond calculating the free energy of T$_{d}$ and 1T$^{\prime}$ \cite{kim_origins_2017, heikes_mechanical_2018}. Assuming the sequence of stacking changes on warming and cooling occur due to becoming energetically preferable, it would be interesting to know how the free energy changes for structures with arbitrary stacking patterns, including for disordered stacking. More generally, it would be good to know what the dominant drivers of stacking changes are, i.e., whether changes in crystal vibrations or band structure is more important. Whatever the source of the effective interlayer interaction, it should be long-range enough to account for the gradual reduction in stacking defects (likely due to annihilating twin boundaries) on cooling into T$_{d}$ or warming into 1T$^{\prime}$ \cite{tao_appearance_2019}.

Even without a microscopic understanding, some insight can be gained by comparing the sliding layer transitions in Mo$_{1-x}$W$_{x}$Te$_{2}$ with those of the chromium trihalides CrX$_{3}$ (X=Cl, Br, I) to that of Mo$_{1-x}$W$_{x}$Te$_{2}$. The chromium trihalides each transition between a high-temperature monoclinic phase and a low-temperature rhombohedral phase along a thermal hysteresis loop \cite{mcguire_crystal_2017}. Each of these phases can be built by a repeated stacking operation, but the operations for the two phases are not symmetry-equivalent with each other (though the outer halide ions of adjacent layers maintain similar positioning), in contrast to the ideally symmetry-equivalent A- and B-type stacking operations of Mo$_{1-x}$W$_{x}$Te$_{2}$. This lack of symmetry-equivalence may explain why a significant discontinuity ($\sim$0.3\%) in the interlayer spacing is reported for CrCl$_{3}$ \cite{mcguire_magnetic_2017} and CrI$_{3}$ \cite{mcguire_coupling_2015}, but not for  Mo$_{1-x}$W$_{x}$Te$_{2}$ in our data (at most $\sim$0.04\% from T$_{d}$ to T$_{d}^{*}$ in MT2.) 
Another similarity between CrX$_{3}$ and MoTe$_{2}$ is the lack of a transition for crystals that are sufficiently thin, with no transition in CrI$_{3}$ for a 35-nm-thick \cite{klein_enhancement_2019} or 4-nm-thick \cite{ubrig_low-temperature_2019} crystal, and no transition in MoTe$_{2}$ for a $\sim$12-nm-thick crystal \cite{he_dimensionality-driven_2018} (and a broadening of the transition with decreasing thickness reported below $\sim$120 nm \cite{cao_barkhausen_2018}.) 
For MoTe$_{2}$, quantum confinement due to thickness reduction has been offered as an explanation for the stability of the T$_{d}$ phase in thin films at room temperature \cite{he_dimensionality-driven_2018}. 
A better understanding of the sliding layer transitions of Mo$_{1-x}$W$_{x}$Te$_{2}$ may improve understanding of the transitions in materials such as CrX$_{3}$ and vice versa.

\section{Conclusion}
\label{sec:Conclusion}

We investigated the structural phase diagram in Mo$_{1-x}$W$_{x}$Te$_{2}$ along hysteresis loops between 1T$^{\prime}$ and T$_{d}$ for $x$ up to $\sim$0.54. With increasing W fraction, the complex behavior first seen in MoTe$_{2}$ transforms into a phase coexistence-like behavior, and the T$_{d}^{*}$ phase is not observed by $x \approx 0.34$. The $\delta$ parameter, and thus the monoclinic $\beta$ angle in the 1T$^{\prime}$ phase, decrease with temperature as well as W. Though a phase-coexistence-like behavior is also observed under pressure, the transition width remains roughly constant under W-substitution, in contrast to the substantial broadening seen under pressure.

\nocite{bish_modern_1989,luo_hall_2015,wu_temperature-induced_2015,zandt_quadratic_2007,rhodes_bulk_2017,cho_te_2017,santos-cottin_low-energy_2020,homes_optical_2015,kimura_optical_2019,paul_controllable_2019,qi_superconductivity_2016}

\section*{Acknowledgements}

This work has been supported by the Department of Energy, Grant number
DE-FG02-01ER45927. A portion of this research used resources at the High Flux Isotope Reactor and the Spallation Neutron Source, which are DOE Office of Science User Facilities operated by Oak Ridge National Laboratory. We acknowledge the support of the National Institute of Standards and Technology, U.S. Department of Commerce, in providing the neutron research facilities used in this work.



%

\end{document}